# Amplitude Contrast Imaging in High-Resolution Transmission Electron Microscopy of Ferroelectric Superlattice Film

Jianguo Wen[1*], Dean J. Miller[1*], Nestor J. Zaluzec[1], Russell E. Cook[1], Ho Nyung Lee[2], Xifan Wu[3]

[1]Electron Microscopy Center, Argonne National Laboratory, Argonne, IL 60439, USA.
[2]Materials Science and Technology Division, Oak Ridge National Laboratory, Oak Ridge, Tennessee 37831, USA.
[3]Department of Physics, Temple University, Philadelphia, PA 19122, USA.

To date, high-resolution electron microscopy has largely relied on using the phase of the exit wave function at the exit surface to form a high-resolution electron microscopic image. We have for the first time used chromatic aberration correction to implement a new imaging mode to achieve amplitude contrast imaging in high-resolution electron microscopy, allowing us to obtain directly interpretable high-resolution electron microscopic images with discrimination between light and heavy atomic columns. Using this imaging approach, we have successfully visualized the atomic structure in a $BaTiO_3/CaTiO_3$ superlattice with high spatial accuracy and discrimination between Ba and Ca columns, providing direct visualization of the Ca and Ba associated oxygen octahedral tilt that controls ferroelectric behavior in these superlattice structures. Furthermore, this approach offers new opportunities to unravel the structure in a wide range of materials, especially complex oxides with exotic behaviors based on specific structural arrangements.

Subject Areas: Materials Science, Strongly Correlated Materials

Interface engineering based on ABO$_3$ perovskite oxide heterostructures has led to new functional properties such as enhanced ferroelectricity [1–4], magnetoelectric coupling [5], spin-structural response [6], and piezoelectricity [7]. Multifunctionality has been achieved through the atomic-level control of the associated BO$_6$ octahedral distortion at interfaces. For example, enhanced piezoelectricity was observed in BaTiO$_3$/CaTiO$_3$ superlatice films [7,8], which was theoretically predicted based on suppression of oxygen octahedral rotations at the interfaces [9]. Understanding the underlying structural basis for these phenomena requires the precise determination of all atomic positions including oxygen at interfaces.

Transmission electron microscopy (TEM) is a powerful tool to visualize atomic structure in real space using high-resolution electron microscopy (HREM). In order to visualize the atomic structure of oxides at interfaces, such as CaTiO$_3$/BaTiO$_3$ ferroelectric superlatice structures, several imaging requirements must be satisfied as follows: 1) distinguish between atomic columns of different species to locate the interface at the atomic scale; 2) observe oxygen columns with a high signal to noise ratio; and 3) provide location spatial accuracy sufficient to resolve oxygen octahedral tilts. Conventional imaging modes generally do not meet all of these requirements.

In a scanning transmission electron microscope (STEM) equipped with a pre-specimen spherical aberration (C$_s$) corrector, high angle annular dark-field (HAADF) imaging mode can provide Z-contrast between atomic columns like Ba and Ca. However, for many materials, it is not possible to image all atomic columns at the same time using HAADF if the material consists of both light and heavy elements, such as oxides [10], and several different approaches may be required. For example, Borisevich *et al.* reported the suppression of octahedral tilts at the BiFeO$_3$-La$_{0.7}$Sr$_{0.3}$MnO$_3$ interface using a combination of STEM imaging modes [11]. This required serial collection of images in different modes. It would be much better if the entire structure could be visualized in a single high quality image.

In a TEM with a post-specimen $C_s$ corrector, Jia *et al.* showed that atomic columns of both heavy and light elements can be observed using an imaging mode termed negative spherical-aberration imaging (NCSI) [12–14]. This mode is achieved by exploiting $C_s$ value to balance contrast and resolution and uses the phase of exit wave to form a HREM image. Although atomic positions are accurately determined using the NCSI imaging mode as shown in Pb(Zr,Ti)O$_3$ films [15,16], all atomic columns including oxygen appear white on a dark background, and as such one cannot readily distinguish between Ba and Ca columns.

In 1998, Foschepoth and Kohl theoretically predicted that amplitude contrast HREM (AC-HREM) can be used to obtain directly interpretable HREM images and distinguish light and heavy atoms in an aberration-corrected TEM [17]. Chen *et al* demonstrated this approach using a $C_s$-corrected TEM (without chromatic aberration ($C_c$) correction). The images they obtained show some features of amplitude contrast but are noisy and not satisfactory for direct structural interpretation [18]. Here, we exploit $C_c$ correction that, together with correction of $C_s$ and other higher order aberrations, allows one to experimentally implement practically useful AC-HREM. This imaging mode allows us obtaining a directly interpretable HREM image of all atomic columns including oxygen columns with spatial accuracy as well as distinguishing light Ca and heavy Ba atomic columns due to channeling contrast. Therefore, all three imaging requirements for the direct visualization of atomic structure at interfaces are satisfied. In this study, we used this new approach to directly observe the suppression of TiO$_6$ octahedral tilt at the interface of BaTiO$_3$ and CaTiO$_3$ in a ferroelectric BaTiO$_3$/CaTiO$_3$ superlattice [19].

As described by channeling theory [20], the phase of a propagating electron wave is a constant over the atomic column and spreads linearly with depth as schematically shown in Fig. 1. This means that phases of adjacent atomic columns overlap quickly when the sample becomes thick, and direct imaging of atomic columns is possible only for a very thin sample. In contrast, the amplitude of the exit wave is strongly peaked at the atomic column position and far less spread with depth as compared to phase. Therefore, the amplitude of the exit wave can be used to accurately determine atomic positions of the projected structure even when the sample is thick. More importantly, amplitude varies periodically with depth as shown in Fig. 1 and the oscillation period, characterized by an extinction distance, is inversely proportional to Z due to different electrostatic potential for different atoms [21]. Thus, imaging using amplitude of the exit wave can provide both good spatial accuracy and discrimination between atomic species based on atomic resolution channeling contrast. As an example, the amplitude and phase of the exit wave as a function of sample thickness for a $CaTiO_3$/$BaTiO_3$ interface is shown in Supplementary Materials S1. Since the amplitude of the exit wave is a function of sample thickness and atomic species within each column, ordered materials with light and heavy atomic columns are expected to show atomic resolution channeling contrast as a function of thickness when the imaging condition is favorable to obtain pronounced amplitude contrast and suppressed phase contrast.

Both conventional or $C_s$ corrected HREM images are dominated by phase contrast as long as $C_c$ is of significant magnitude. When both $C_s$ and $C_c$ are uncorrected in a conventional TEM, HREM images are taken under the Scherzer defocus to maximize phase contrast and resolution. [22] For a TEM with a $C_s$ image corrector, correction of $C_s$ towards zero improves resolution, but phase contrast decreases dramatically due to the limit of the uncorrected $C_c$ damping envelope as shown in the Supplementary Materials S2. To balance resolution and phase contrast, Jia *et al* introduced the NCSI method in which a negative $C_s$ value (~ -10 to ~ -40 μm) and a corresponding Lichte defocus are chosen to gain strong phase contrast with sufficient resolution to facilitate direct structural mapping [12,23,24]. Under typical NCSI imaging conditions, HREM images are dominated by phase contrast as shown in Supplementary Materials S3.

$C_c$ correction allows correction of $C_s$ towards zero to improve resolution *without* compromising contrast. Thus, when both $C_s$ and $C_c$ are corrected to small values, sub-Angstrom HREM using amplitude can be realized. As shown in Fig. 2, $C_c$ correction extends the $C_c$ damping envelope significantly towards high spatial frequencies. In this case, both the magnitude and the resolution for phase contrast and amplitude contrast increase without compromising contrast. Under perfect aberration correction, i.e. zero defocus $\Delta f$ (or $C_1$), zero spherical aberration $C_s$ (or $C_3$), and ignoring higher order aberrations ($C_5, C_7$ …), the phase contrast of a thin object vanishes, and the remaining contrast is only amplitude contrast [25]. For practical situations, we cannot achieve such perfect imaging conditions, but the closer one gets to perfect aberration correction, the more dominant the amplitude contrast component becomes. For example, when $C_c$ is uncorrected, HREM images obtained using nearly zero $C_s$ and $\Delta f$ contain amplitude contrast as shown in Fig. 3a (similar to Fig. 5a in the ref [18]), but they are noisy, exhibit more residual phase contrast, and do not provide sufficient resolution and contrast for direct structural interpretation. However, for a TEM with both $C_s$ and $C_c$ image corrector, under practically achievable amplitude contrast imaging (ACI) conditions (i.e. small but non-zero $C_s$ and $C_c$ values), simulated HREM images are strongly dominated by amplitude contrast as shown in Supplementary Materials S4. For example, the filled blue curves in Figs. 2a and 2b show phase contrast-transfer function (PCTF) and amplitude contrast-transfer function (ACTF) under an ACI imaging condition ($\Delta f = 1$ nm, $C_s = -3$ μm, and $C_c = 0.04$ mm). Under this ACI imaging condition, the ACTF has a value of near unity up to the information limit of the instrument, while the corresponding PCTF is suppressed to low values out to the information limit of the instrument. In essence, this condition approaches the perfect correction of aberration resulting in maximum amplitude contrast and minimum phase contrast [25]. This is demonstrated in Fig. 3b, which shows an AC-HREM image of Si [110] under an ACI imaging condition. It clearly shows that $C_c$ correction allows obtaining AC-HREM with good resolution and good contrast. In addition to enhancing the amplitude contrast and suppressing the phase contrast, $C_s$ and $C_c$ correction also reduces image delocalization as shown in Fig. 2c. More details can be found in Supplementary Materials S5.

To illustrate the contribution of $C_c$ correction to AC-HREM further, we use a $CaTiO_3/BaTiO_3$ interface as an example to compare the calculated phase and amplitude images of the exit wave with simulated HREM images as shown in Fig. 4. Fig. 4b and 4c show the calculated phase and amplitude of the exit wave for a 9 nm thick sample of a $CaTiO_3/BaTiO_3$ interface as schematically shown in Fig. 4a. Fig. 4d shows a simulated HREM image under an NCSI condition calculated using JEMS [26]. Calculated cross-correlation coefficients between the simulated image (d) with the phase (b) and amplitude (c) of exit wave are 0.641 and 0.585 respectively, indicating that HREM images obtained under medium $C_s$ values exhibit significant phase contrast. Even when $C_s$ is reduced to a lower value such as $C_s = -3$ μm, the simulated image shown in (e) more closely resembles the phase image in (b). Fig. 4f shows the simulated image under an ACI condition under the corresponding Scherzer defocus. The measured cross-correlation coefficients between the simulated image (4f) with the phase (4b) and amplitude (4c) of the exit wave are 0.417 and 0.865 respectively, indicating that HREM images obtained when both $C_s$ and $C_c$ are corrected are dominated by amplitude contrast.

AC-HREM provides several major advantages over phase contrast HREM as discussed by Lentzen [27]. First, an AC-HREM image has one-to-one correspondence to the projection of atomic columns with good spatial accuracy since the amplitude of exit wave is strongly peaked at each atomic column position. This one-to-one correspondence is similar to HAADF. Since image delocalization is nearly zero (see Fig. 2c), the image directly represents the object through the amplitude of exit wave. This enables one to obtain accurate projected positions for all atomic columns including defects. Second, AC-HREM provides direct structural mapping even for thicker samples due to the strong channeling effect. Simulations indicate direct structural mapping can be obtained for a sample up to 20 nm in thickness for a $BaTiO_3/CaTiO_3$ interface along [110] over a wide range of defocus. Third, an amplitude contrast HREM image contains channeling contrast for each atomic column [20]. For a thin sample, the amplitude contrast between different atomic columns is not strong, and a phase contrast HREM imaging mode like NCSI is more useful to enhance contrast. However, with increasing sample thickness, the amplitude intensity for each column becomes more pronounced, and an AC-HREM imaging mode like ACI can be more useful. By choosing a suitable sample thickness and defocus, one can reach an imaging condition that yields a contrast between different atomic columns based on the modulation of channeling contrast. For example, the extinction distance of Ba and Ca columns in $BaTiO_3$ and $CaTiO_3$ along the [110] zone axis in our $ABO_3$ ferroelectric heterostructure is about 8 nm and 14 nm, respectively, as estimated from simulated images. Thus, when the specimen thickness is around 8 nm to 10 nm, the exit waves of Ba and Ca columns are expected to show opposite contrast, i.e. when the Ca columns are at maximum amplitude, the Ba columns are at minimum amplitude.

Fig. 5 shows the value of the AC-HREM imaging approach in correlating the structure of a $(BaTiO_3)_4/(CaTiO_3)_4$ ferroelectric superlattice to its properties [28]. In this HREM image, channeling contrast between Ba and Ca is clearly observed, in which atomic columns of CaO and BaO appear bright and dark, respectively. Oxygen and Ti columns appear as bright. The overall structure of alternating $BaTiO_3$ and $CaTiO_3$ is directly visualized using this imaging approach. More importantly, AC-HREM provides a direct and accurate measurement of the positions of oxygen columns that allows us to quantify $TiO_6$ tilt angles (along the [110] direction) as a function of distance from the interface (More details can be found in Supplementary Materials S6). The boxed columns marked 1,2,3 in Fig. 5a reveal the differences of $TiO_6$ octahedra orientation at the interface compared to that within the $BaTiO_3$ and $CaTiO_3$ sublayers. Oxygen octahedra in the $BaTiO_3$ sublayer do not tilt as indicated by flat $TiO_2$ planes (box 1) between adjacent BaO layers. In contrast, oxygen octahedra in the $CaTiO_3$ sublayer are tilted by ~ 9 degrees along the [110] direction as indicated by corrugated $TiO_2$ planes (box 2) between adjacent CaO layers. When a $TiO_2$ plane (box 3) is sandwiched between a BaO and a CaO layer at the interface, oxygen octahedral tilt is nearly suppressed (average tilt angle ~ 3 degrees). This interface effect is consistent with the predicted suppression of octahedral rotation (along the [001] direction) at the $BaTiO_3/CaTiO_3$ interface based on density functional theory (DFT) calculations [9].

Importantly, AC-HREM now allows us to reveal mixing between Ba and Ca at the interface (dashed box region of Fig. 5a) by a subtle change in the contrast of individual columns. This mixing can be measured quantitatively by comparing the intensity profile of the Ca/Ba plane illustrated in Fig. 5b with the JEMS calculated image shown in Fig. 5c. Simulation indicates that the mixing ratio at columns 1, 2, 3, 4, and 5 are $Ca_{0.4}Ba_{0.6}$, $Ca_{0.5}Ba_{0.5}$, $Ca_{0.7}Ba_{0.3}$, $Ca_{0.7}Ba_{0.3}$, and $Ca_{0.6}Ba_{0.4}$ respectively. Since the simulation indicates a sample thickness of 9 nm corresponding to 16 cation atoms per column, these mixing ratios correspond to atom ratios Ca/Ba of 6/10, 8/8, 11/5, 10/6, for columns 1, 2, 3, 4, and 5 respectively. Thus, the distinguishable contrast between these columns suggests the sensitivity of the technique is approximately two atoms for this sample under these imaging conditions. This result is close to the prediction of single atom sensitivity suggested by Aert *et al* and Wang *et al* [20,29,30]. Detailed image simulation suggests a lower limit of detection of Ca occupancy around 0.3. In addition, the oxygen octahedral tilt angle right below the columns 1, 2, 3, 4, and 5 gradually increases with increasing mixing ratio of Ca/Ba. This indicates $TiO_6$ tilt is strongly localized both parallel and perpendicular to [001] direction, suggesting $TiO_6$ tilt is mainly determined by the nearest neighbor chemical environment. As shown in box 4 of Fig. 5, the presence of even one CaO column in the BaO layer is enough to cause a strong oxygen octahedral tilt.

In summary, amplitude contrast high-resolution electron microscopy provides directly interpretable atomic scale elemental and structural information in a single high-resolution electron microscopic image. Using this imaging approach, we have successfully visualized the A-site associated oxygen octahedral tilt in the ferroelectric $BaTiO_3$/$CaTiO_3$ superlattice. We found that the suppression of oxygen octahedral tilt by Ba atoms is localized at atomic scale.

The electron microscopy was accomplished at the Electron Microscopy Center at Argonne National Laboratory, a U.S. Department of Energy Office of Science Laboratory operated under Contract No. DE-AC02-06CH11357 by UChicago Argonne, LLC. X.W. thanks the support from Air Force Office of Scientific Research with award no. FA9550-13-1-0124. The work at Oak

Ridge National Laboratory was supported by the U.S. Department of Energy, Basic Energy Sciences, Material Sciences and Engineering Division.

[*]Corresponding authors: jgwen@anl.gov, miller@anl.gov

**References:**

[1]  E. Bousquet, M. Dawber, N. Stucki, C. Lichtensteiger, P. Hermet, S. Gariglio, J.-M. Triscone, and P. Ghosez, Nature **452**, 732 (2008).
[2]  H. N. Lee, H. M. Christen, M. F. Chisholm, C. M. Rouleau, and D. H. Lowndes, Nature **433**, 395 (2005).
[3]  M. Stengel, D. Vanderbilt, and N. A. Spaldin, Nat. Mater. **8**, 392 (2009).
[4]  R. Cohen, Nat. Mater. **8**, 366 (2009).
[5]  R. Ramesh and N. A. Spaldin, Nat. Mater. **6**, 21 (2007).
[6]  J. H. Lee and K. M. Rabe, Phys. Rev. Lett. **104**, 207204 (2010).
[7]  J. Y. Jo, R. J. Sichel, H. N. Lee, S. M. Nakhmanson, E. M. Dufresne, and P. G. Evans, Phys. Rev. Lett. **104**, 207601 (2010).
[8]  S. S. A. Seo, J. H. Lee, H. N. Lee, M. F. Chisholm, W. S. Choi, D. J. Kim, J. Y. Jo, H. Kim, J. Yu, and T. W. Noh, Eprint ArXiv07091791 (2007).
[9]  X. Wu, K. M. Rabe, and D. Vanderbilt, Phys. Rev. B **83**, 020104 (2011).
[10]  A. Ohtomo, D. A. Muller, J. L. Grazul, and H. Y. Hwang, Nature **419**, 378 (2002).
[11]  A. Y. Borisevich, H. J. Chang, M. Huijben, M. P. Oxley, S. Okamoto, M. K. Niranjan, J. D. Burton, E. Y. Tsymbal, Y. H. Chu, P. Yu, R. Ramesh, S. V. Kalinin, and S. J. Pennycook, Phys. Rev. Lett. **105**, 087204 (2010).
[12]  C. L. Jia, M. Lentzen, and K. Urban, Science **299**, 870 (2003).
[13]  C. L. Jia and K. Urban, Science **303**, 2001 (2004).
[14]  K. W. Urban, C.-L. Jia, L. Houben, M. Lentzen, S.-B. Mi, and K. Tillmann, Philos. Trans. R. Soc. Math. Phys. Eng. Sci. **367**, 3735 (2009).
[15]  C.-L. Jia, S.-B. Mi, K. Urban, I. Vrejoiu, M. Alexe, and D. Hesse, Nat. Mater. **7**, 57 (2007).
[16]  C.-L. Jia, K. W. Urban, M. Alexe, D. Hesse, and I. Vrejoiu, Science **331**, 1420 (2011).
[17]  M. Foschepoth and H. Kohl, Phys. Status Solidi A **166**, 357 (1998).
[18]  J. H. Chen, H. W. Zandbergen, and D. V. Dyck, Ultramicroscopy **98**, 81 (2004).
[19]  S. S. A. Seo and H. N. Lee, Appl. Phys. Lett. **94**, 232904 (2009).
[20]  A. Wang, F. R. Chen, S. Van Aert, and D. Van Dyck, Ultramicroscopy **110**, 527 (2010).
[21]  D. Van Dyck and M. Op de Beeck, Ultramicroscopy **64**, 99 (1996).
[22]  O. Scherzer, J. Appl. Phys. **20**, 20 (1949).
[23]  H. Lichte, Ultramicroscopy **38**, 13 (1991).
[24]  E. Spiecker, M. Garbrecht, W. Jäger, and K. Tillmann, J. Microsc. **237**, 341 (2010).
[25]  M. Lentzen, C. Jia, and K. Urban, Microsc. Microanal. **9**, 48 (2003).
[26]  Stadelmann, P. Java EMS http://cimewww.epfl.ch/people/stadelmann/jemsWebSite/jems.html.
[27]  M. Lentzen, A. Thust, and K. Urban, Microsc. Microanal. Off. J. Microsc. Soc. Am. Microbeam Anal. Soc. Microsc. Soc. Can. **10**, 980 (2004).
[28] TEM sample preparation: The TEM sample was prepared using a focused-ion beam (FIB). The superlattice film was first coated with a thin layer of carbon using e-beam deposition to protect the film from ion beam damage. A thin section of the superlattice film on an $SrTiO_3$ substrate was lifted off and mounted on a FIB TEM grid. The thin section was thinned and polished with 30 kV Ga+ ions with a 50 pA beam current. The FIB prepared TEM sample was further polished with 500 eV Ar+ ions with 15     A of current to remove the damaged layer induced by the Ga+ ion beam.

[29]  S. Van Aert, P. Geuens, D. Van Dyck, C. Kisielowski, and J. R. Jinschek, Ultramicroscopy **107**, 551 (2007).
[30]  A. Wang, F. R. Chen, S. Van Aert, and D. Van Dyck, Ultramicroscopy **116**, 77 (2012).

**Figure Captions**

**Fig. 1**. Schematic representation of phase and amplitude of electron channeling through atomic columns with different atomic numbers. Phase expands linearly but amplitude oscillates around columns with depth. This figure demonstrates how imaging using the amplitude of exit wave can provide discrimination between light and heavy atomic columns.

**Fig. 2**. Contribution of $C_c$ correction to achieve amplitude contrast HREM (AC-HREM). (a) Phase contrast transfer function (PCTF) for two imaging conditions: NCSI (green, $C_s$ = -40 µm, $\Delta f$ = 8 nm, $C_5$=2 mm, $C_c$ = 1 mm) and AC-HREM (blue, $C_s$ = -3 µm, $\Delta f$ = 1 nm, $C_5$=2 mm, $C_c$ = 0.04 mm). (b) Amplitude contrast transfer function (ACTF) for NCSI (green) and AC-HREM (blue) imaging conditions. (c) Image delocalization for the AC-HREM imaging condition. The vertical dashed line in each figure indicates the information limit of our instrument. The red curves in (a) and (b) indicate how $C_c$ correction extends the $C_c$ damping envelope from the dashed curve ($C_c$ = 1.0 mm) to the solid curve ($C_c$ = 0.04 mm) with high spatial frequencies, enabling AC-HREM (blue), while at the same time image delocalization is reduced (c).

**Fig. 3**. Amplitude contrast HREM image of Si [110] for a) uncorrected $C_c$ ($C_c$ = 1.9 mm, $C_s$ < 0.5 µm), b) corrected $C_c$ ($C_c$ = 1 mm, $C_c$ < 0.5 µm) demonstrating the effect of $C_c$ correction in improving resolution and contrast. The images were acquired on adjacent areas of the sample using identical imaging conditions except of the value of $C_c$. c) and d) show intensity profiles at the lines indicated in a) and b). White dots surrounded by a dark ring in b) are the typical features caused by channeling contrast.

**Fig. 4**. Simulated images of the $CaTiO_3/BaTiO_3$ interface demonstrating that $C_c$ correction enables AC-HREM. (a) Atomic arrangement along [110]. (b) Calculated exit wave phase image and (c) calculated exit wave amplitude image for a 9 nm thick sample. (d-g) Simulated HREM images. For an uncorrected $C_c$ of 1.5mm (d, e), simulated images using (d) a medium $C_s$ value ($C_s$ = -40 μm) and (e) a small $C_s$ value ($C_s$ = -3 μm) show close correlation to phase image of the exit wave in (b). For a corrected $C_c$ of 0.1 mm and a small $C_s$ value of -3 μm (f), the simulated image shows a close correlation to the amplitude image of the exit wave in (c). Adjusting defocus from (f) Δf = 3 nm to (g) Δf = 1 nm improves the amplitude contrast and reduces image delocalization further. The arrows indicate atomic columns with $Ca_{0.5}Ba_{0.5}$. $C_5$ = 2 mm in all simulated images. Simulation software: JEMS, multislice method.

**Fig. 5**. Experimental HREM image demonstrating AC-HREM for a $CaTiO_3/BaTiO_3$ superlattice film. (a) HREM image of a $CaTiO_3/BaTiO_3$ superlattice along [110] under an AC-HREM imaging condition ($C_s$ = 3 μm, Δf = -2 nm, $C_5$ = -2 mm, $C_c$ = 1 μm, as measured using the Zemlin Tableau method using an amorphous carbon film). Atom columns of CaO, Ti, and O in this image appear as bright dots while atom columns of BaO appear dark. The $TiO_2$ planes are flat when between two BaO layers (box 1) and corrugated when between two CaO layers (box 2). The presence of BaO on one side of a $TiO_2$ plane results in a suppression of oxygen octahedral rotation (box 3). At the interface, the presence of even *one* CaO column in the BaO layer is enough to cause a strong oxygen octahedral tilt (box 4). (b) Magnified image from the dashed box (box 5) in (a) showing the atomic-resolution channeling contrast for CaO/BaO columns. (c) Simulated image indicates the atomic-resolution channeling contrast is sensitive to Ca/Ba occupancy ratio. Columns 1, 2, 3, 4, 5 for the simulated structure are $Ca_{0.4}Ba_{0.6}$, $Ca_{0.5}Ba_{0.5}$, $Ca_{0.7}Ba_{0.3}$, $Ca_{0.7}Ba_{0.3}$, and $Ca_{0.6}Ba_{0.4}$ respectively. The intensity profile from experimental image in (b) matches well with the simulated intensity profile. Note that the brighter CaO/BaO column in the dashed box in (b), the larger tilt angle of $TiO_6$ octahedral underneath each CaO/BaO column.

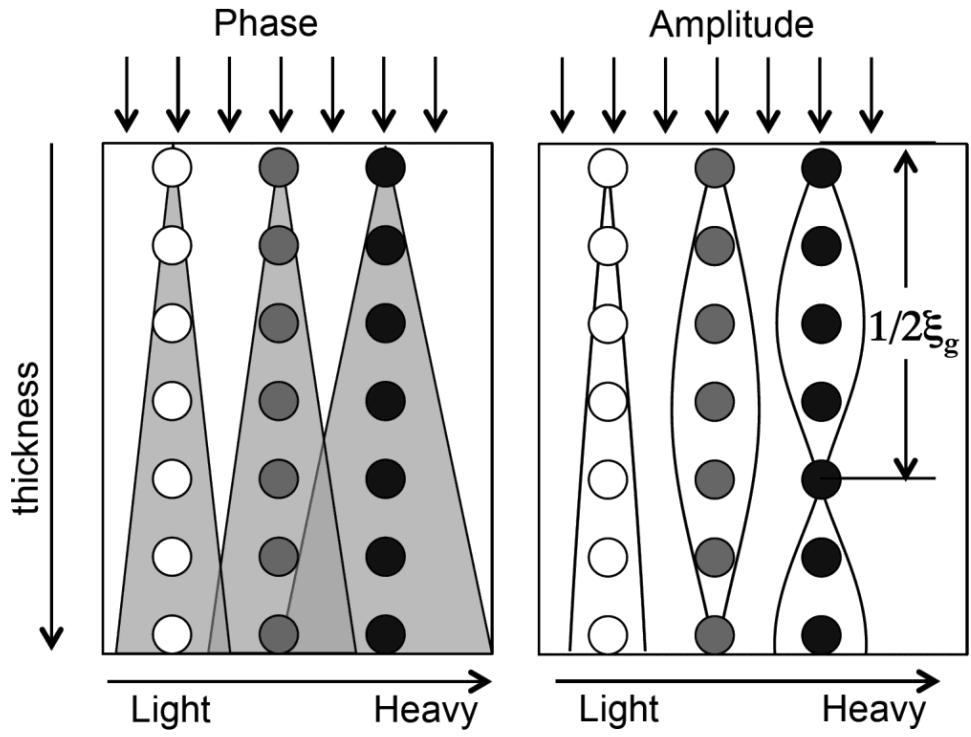

Fig. 1

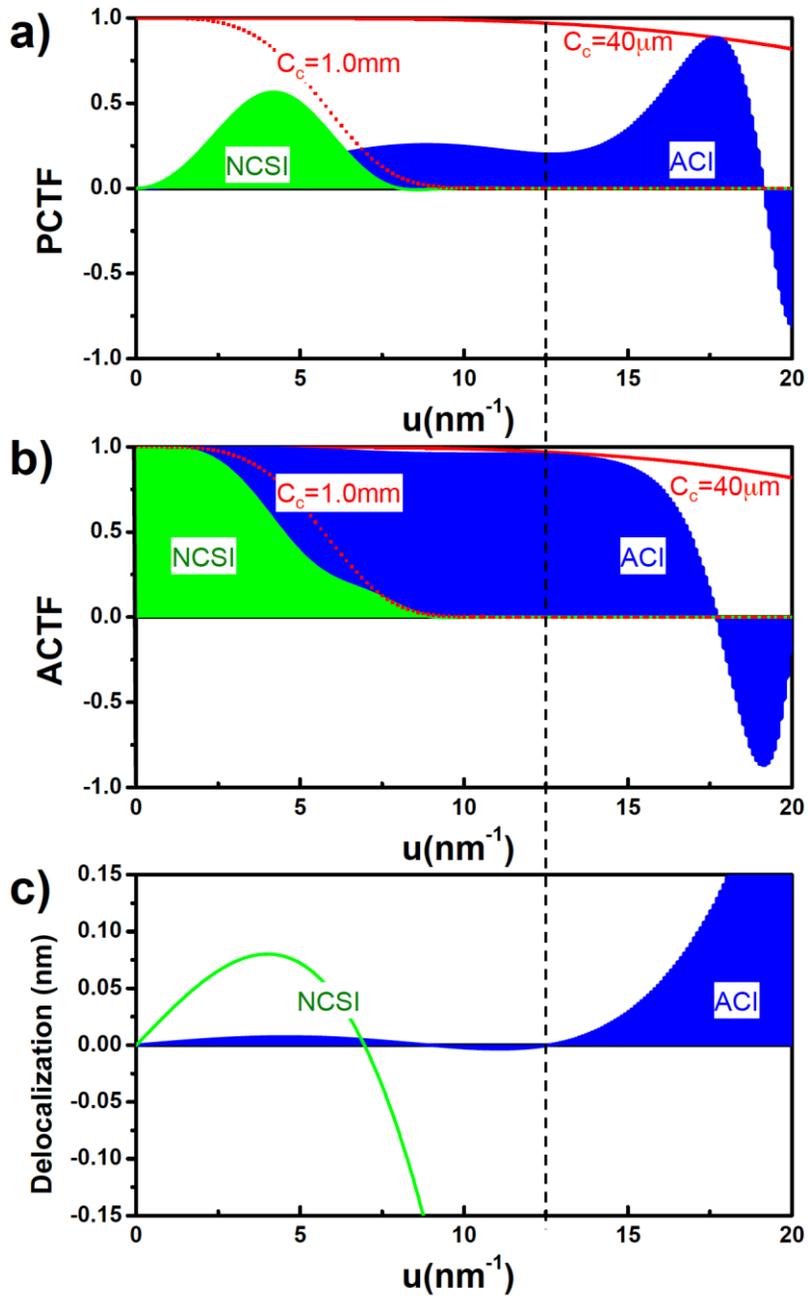

Fig. 2

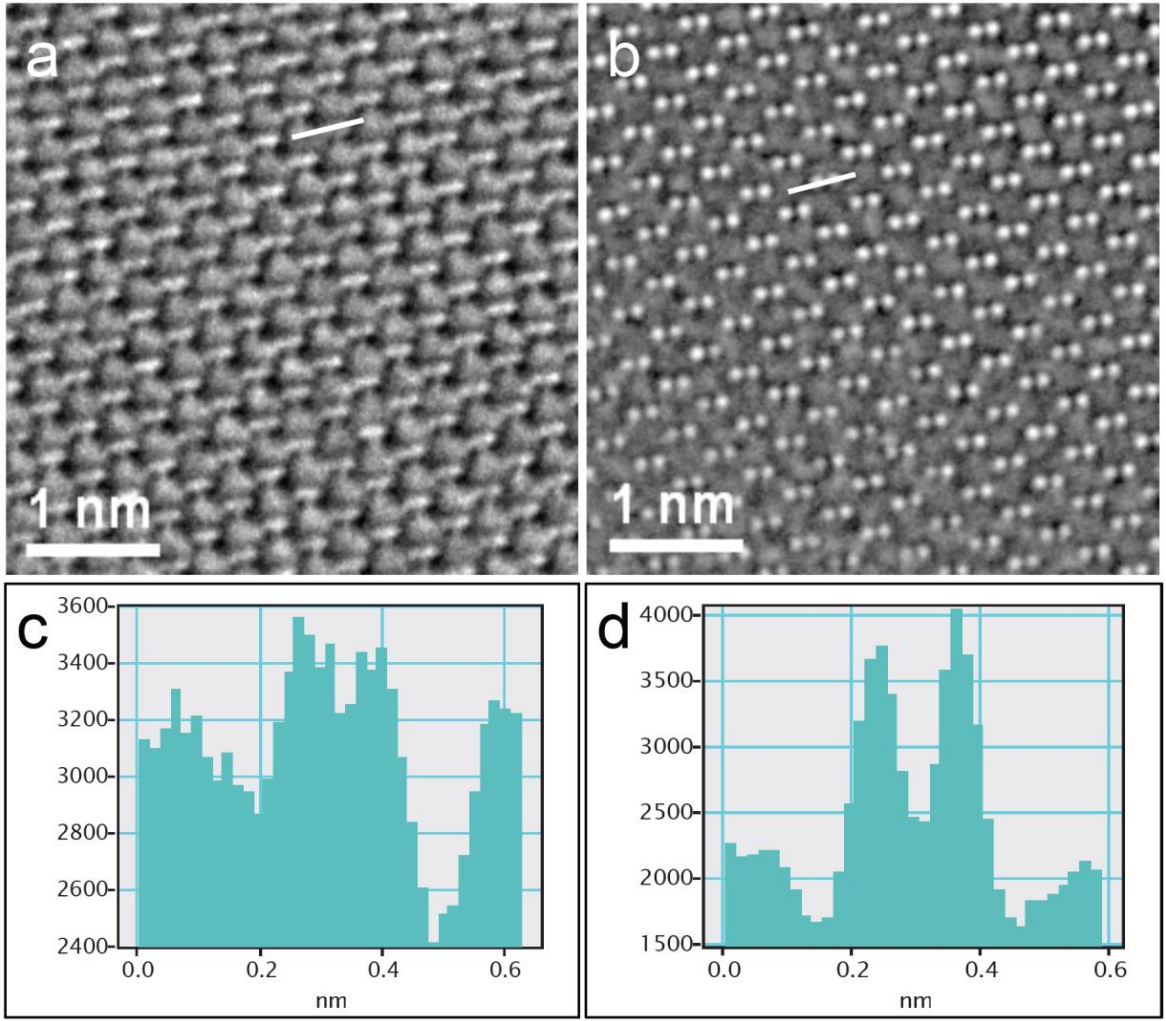

Fig. 3

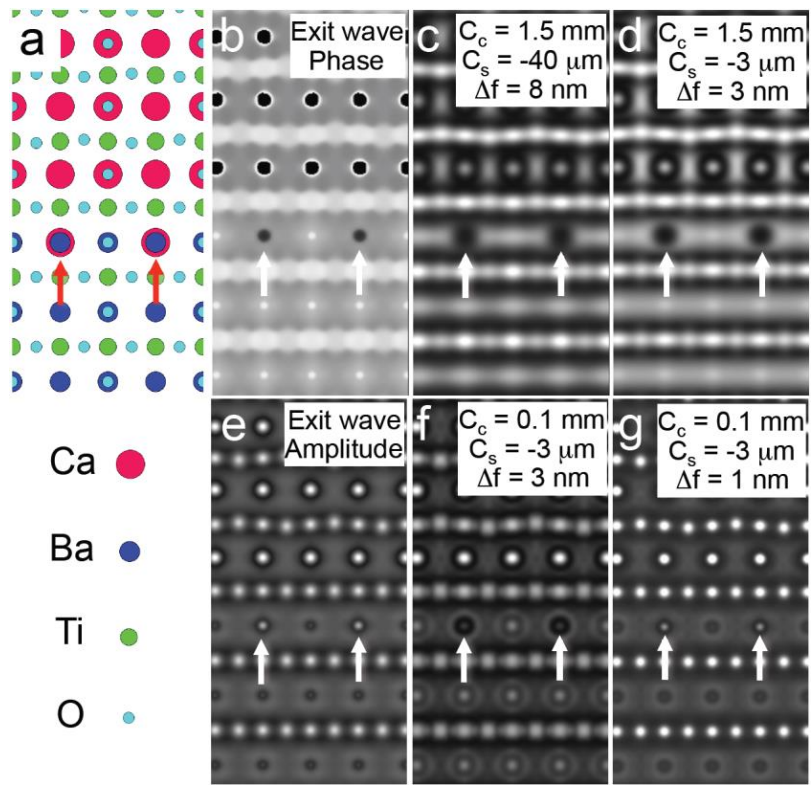

Fig. 4

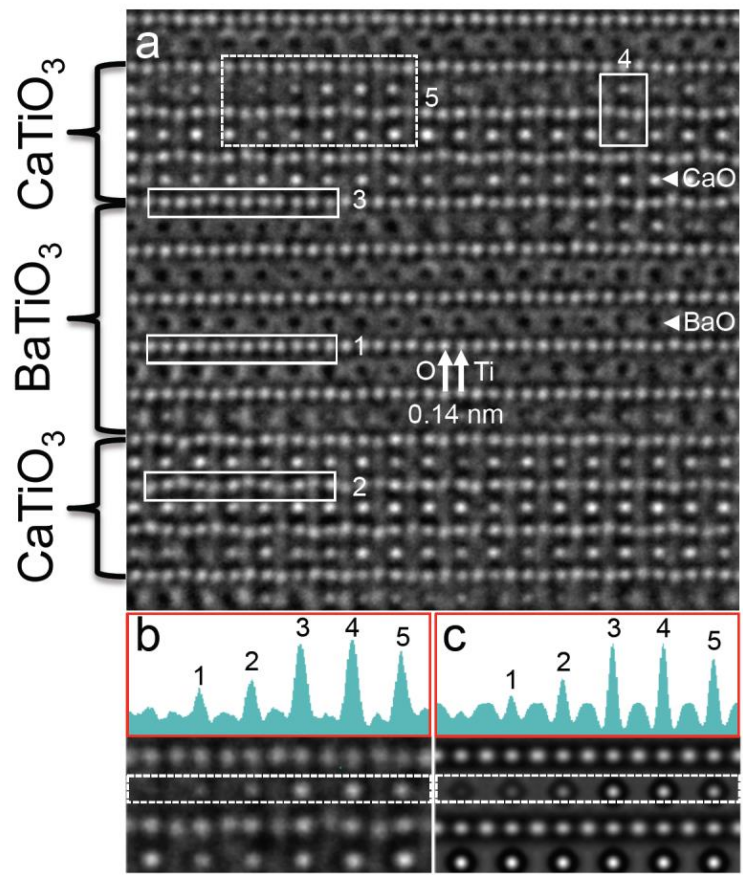

Fig. 5